\documentclass[aps,prd,twocolumn,showkeys,showpacs,longbibliography,superscriptaddress,groupedaddress,amsfonts]{revtex4-1}
\usepackage{graphicx}  
\usepackage{dcolumn}   
\usepackage{bm} 
\usepackage{latexsym}      
\usepackage{amssymb}   
\usepackage{mathrsfs}
\usepackage{amsmath}
\usepackage{txfonts}
\usepackage{multirow}
\usepackage{hyperref}
\usepackage{latexsym}
\usepackage{graphics} 
\usepackage{dcolumn}   
\usepackage{bm}        
\usepackage{amssymb}   
\usepackage{mathrsfs}
\usepackage{amsmath}

\newcommand{\RomanNumeralCaps}[1]
{\MakeUppercase{\romannumeral #1}}
\usepackage{enumerate}

\begin{document}
\title{Dynamical system analysis of Quintom Dark Energy Model }
\author{Sudip Mishra \thanks{corresponding author}}
\email[]{sudipcmiiitmath@gmail.com}
\author{Subenoy Chakraborty}
\email[]{schackraborty.math@gmail.com}
\affiliation{ Department of Mathematics, Jadavpur University, Kolkata- 700032, WB, India.}




\maketitle

\section{Introduction}
An important problem in present cosmology is to comprehend the role of dark energy (DE) which discovered at the turn of last century when two independent observational studies \cite{0803.0547[astro-ph],astro-ph/1701.08165} from Type Ia Supernovae (SNIa) \cite{astro-ph/0404062,astro-ph/9805201} revealed that the universe is going through cosmic acceleration at a fast pace.  The other two important evidences to support of the role of DE \cite{astro-ph/0402265, *gr-qc/0405038, *gr-qc/0410050} are based on the experimental study of  cosmic microwave background radiation along with large-scale structure surveys (CMB $\&$ LSS).  The salient quantity of DE is its equation of state (EoS) which explicitly defined as $\omega_{DE}=\frac{p_{DE}}{\rho_{DE}}$  where $p_{DE}$ and $\rho_{DE}$ are the pressure and energy densities respectively.  If we restrict ourselves in four dimensional Einstein’s gravity, nearly all DE models can be classified by the behaviors of equations of state (EoS).  For example, the case of a non-zero and positive cosmological constant boundary corresponds to $\omega_{\Lambda}=-1$.  In this case, $\rho_{\Lambda}$ is independent of the scale factor a(t).  A quintessence field is dynamical field for which the barotropic parameter of the dark energy equation of state is above the $\Lambda$CDM boundary \cite{1205.3421[gr-qc],*hep-th/0212290}, that is $\omega_Q>-1$.  Similarly for a phantom field \cite{hep-ph/0311070, *astro-ph/0405518}, $\omega_p<-1$. Interestingly some data analyses suggest the cosmological constant boundary (or phantom divide) is crossed \cite{astro-ph/0311293,*astro-ph/0401293,*astro-ph/0407421,astro-ph/0404540, gr-qc/0406098,*0408225}, due to the dynamical behavior of the dark energy EoS \cite{hep-th/0603057, *hep-th/1207.6691}.  Moreover, the quintessence and phantom models alone cannot explain the evolution of the dark energy equation of state and the possible crossing of the phantom divide line.\par

According to the Null Energy Condition (NEC), the EoS of normal matter should not be smaller than the cosmological constant boundary.  On the other hand, there exists a ``no-go theorem'' \cite{1803.11084[hep-th],1802.09155[gr-qc], 1607.02047[astro-ph.HE], 1805.03899[gr-qc], *Lipkin}  that prevents the EoS of a single scalar field to cross over the cosmological constant boundary. One possible solution to this problem is to introduce a superposition between two dynamical scalar fields- i.e., a canonical field $\phi$ and a phantom field $\sigma$. Such phenomenological models are known as quintom models which give rise to quintom cosmology  \cite{0909.2776[hep-th], *astro-ph/0410654, *0810.1427[hep-th]}.  Curiously, some of the recent observational data show a significant accordance with a dynamical EoS for the dark energy component corresponding to quintom models.  In these models, the dark energy equation of state parameter presenting an evolution from a phantom behavior $\omega_p<-1$ around present epoch, towards a quintessence behavior $\omega_Q >-1$ in the near past\cite{hep-th/0804.0553,astro-ph/0810.4775,hep-th/0807.3807}.  In this regard we need to mention that a dynamically valid dark energy quintom model requires to have at least two degrees of freedom\cite{0908.2362[astro-ph.CO]}.\par

The present work is related to quintom dark energy cosmological model.  Due to non-linear coupled system of field equations analytic cosmological solutions are not possible.  So dynamical system analysis \cite{1503.05750[gr-qc], *1604.07636[gr-qc]} has been discussed here.  The plan of the present work is as follows:  The basic equations for the quintom cosmological model has been presented in section \RomanNumeralCaps{2}.  Autonomous system has been formed and stability analysis of the line of critical points has been discussed in section \RomanNumeralCaps{3}.  Also bifurcation scenarios have been examined in this section.  The paper ends with a brief discussion on cosmological implications of dynamical system analysis in section \RomanNumeralCaps{4}. 

\section{Basic Equations}
This section is devoted to the basic equations related to quintom model.  Here gravity is minimally coupled to a normal scalar field $\phi$ and a phantom (i.e. negative kinetic energy) scalar field $\sigma$ with a coupled potential $v(\phi,\sigma)$.  The action for this model is described by 
\begin{equation}
S=\int d^4x\sqrt{-g}[\frac{R}{2k^2}+\frac{1}{2}g^{\mu \gamma}\partial_\mu \phi \partial\gamma \phi-\frac{1}{2}g^{\mu \gamma}\partial_\mu \sigma \partial_\gamma \sigma + v(\phi,\sigma)+ \mathcal{L}_m]
\label{IntMod}
\end{equation}
where $k^2=8\pi G$ is the gravitational coupling and $\mathcal{L}_m$ represents the Lagrangian density of matter fields.  In the background of the homogeneous and isotropic flat Friedmann-Lemaitre-Robertson-Walker (FLRW) space-time, the line element is given by
\begin{equation}
ds^2=-dt^2+a^2(t)[dx^2+dy^2+dz^2].
\label{LinEle}
\end{equation}
The explicit form of the Lagrangian \cite{1607.03396[gr-qc]} is
\begin{equation}
L(a,\dot{a},\phi,\dot{\phi},\sigma,\dot{\sigma})=-3a\dot{a}^2 + a^3(\frac{1}{2}\dot{\phi}^2 -\frac{1}{2}\dot{\sigma}^2 + v(\phi,\sigma))
\label{lagrangian}
\end{equation}
Hence `a(t)' is the usual scale factor, $\phi=\phi(t)$, $\sigma=\sigma(t)$ are the canonical and non-canonical scalar fields and an over dot represents differentiation with respect to the cosmic time t.\\
Now varying the action with respect to the scale factor `a(t)' (assuming $\mathcal{L}_m=0$) gives the two Friedmann equations
\begin{equation}
3\frac{\dot{a}^2}{a^2}=\frac{1}{2}\dot{\phi}^2-\frac{1}{2}\dot{\sigma}^2+v(\phi,\sigma)
\label{EinFld1}
\end{equation}
and
\begin{equation}
2\frac{\ddot{a}}{a}+\frac{\dot{a}^2}{a^2}=-\frac{1}{2}\dot{\phi}^2 + \frac{1}{2}\dot{\sigma}^2 + v(\phi,\sigma),
\label{EinFld2}
\end{equation}
while variation of the action w.r.t. the scalar fields give their evolution equations as
\begin{equation}
\ddot{\phi}+3H\dot{\phi}+\frac{\partial v}{\partial \phi}=0
\label{EvoEqn1}
\end{equation}
and
\begin{equation}
\ddot{\sigma}+3H\dot{\sigma}-\frac{\partial v}{\partial \sigma}=0,
\label{EvoEqn2}
\end{equation}
where $H=\frac{\dot{a}}{a}$ is the usual Hubble parameter.  The last two equations (i.e. equations (\ref{EvoEqn1} and \ref{EvoEqn2}) are also known as energy conservation equations for the scalar fields.\\
Now combining the two fluids as a single matter part, the effective equation of state can be written as 
\begin{equation}
\omega_{eff}=\frac{\dot{\phi}^2-\dot{\sigma}^2-2v(\phi,\sigma)}{\dot{\phi}^2-\dot{\sigma}^2+2v(\phi,\sigma)},
\label{EoS}
\end{equation}
It should be noted that the present effective cosmological model will be characterized as quintessence model if $\omega_{eff}\geqslant-1$ i.e. $\dot{\phi}\geqslant \dot{\sigma}$ while it will be phantom in nature if $\omega_{eff}<-1$ i.e. $\dot{\phi}<\dot{\sigma}$.  As we have not assumed any direct coupling between the two scalar fields so in the present work two possible choices of the potential function are considered namely,
\begin{enumerate}[i.]
	\item $v(\phi,\sigma) = v_0(\phi^n-\sigma^m) + \mu_0$, $(\mu_0 ~\textrm{is an arbitrary constant} >0)$\\
	\item $v(\phi,\sigma)=e^{-\lambda(\phi + \sigma)}$
\end{enumerate}
where $v_0$ and $\lambda$ dimensionless constants characterizing the slop of the potential (as defined in section (\ref{StabilityAnalysis})).\\
It is desirable to have $w_{eff}$ $(\neq 0)$ close to the cosmological constant boundary (i.e $w_{eff}=-1$) for a feasible quintom model and as a result, the dynamical evolution
of both scalar fields results a quintom scenario with a smooth transition across $w_{eff}=-1$.  Now differentiating equation (\ref{EoS}) and using the scalar field evolution equations (\ref{EvoEqn1} and \ref{EvoEqn2}) one gets
\begin{equation}
\frac{dw_{eff}}{dt}=\frac{-24vH(\dot{\phi}^2-\dot{\sigma}^2)-4(\dot{\phi}^2-\dot{\sigma}^2+2v)(\dot{\phi}v_{\phi}+\dot{\sigma}v_{\phi})}{(\dot{\phi}^2-\dot{\sigma}^2+2v)^2}
\end{equation}
This equation used as a consistency check when any closed form solution is not possible.  Lastly, in cosmology the expansion of the universe is characterized by the deceleration parameter $q=-(1+\frac{\dot{H}}{H^2})$ with $q<0$ indicating accelerated expansion while $q>0$ indicating decelerated expansion.

\section{Stability of critical points and Bifurcation analysis}\label{StabilityAnalysis}

\subsection{ $v(\phi,\sigma) = v_0(\phi^n-\sigma^m) + \mu_0$} 

In order to reveal the autonomous structure of the cosmological dynamical system described by equations (\ref{EinFld1})-(\ref{EvoEqn2}), we introduce the following variables,  $\phi=y$, $\sigma=z$, $\dot{y}=r$, $\dot{z}=s$.  Thus the Einstein field equations (\ref{EinFld1}, \ref{EinFld2}) and evolution equations (\ref{EvoEqn1}, \ref{EvoEqn2}) turn into an autonomous system as follows
\begin{eqnarray}
\dot{H}&=-\frac{1}{2}r^2+\frac{1}{2}s^2  \label{auto1a}\\
\dot{y}&=r  \label{auto1b}\\
\dot{r}&=-3Hr-nv_0y^{n-1}  \label{auto1c}\\
\dot{z}&=s  \label{auto1d}\\
\dot{s}&=-3Hs-mv_0z^{m-1}  \label{auto1e}
\end{eqnarray}
The over dot represents the differentiation with respect to ` t ' and m, n are choosing to be positive integer greater than 1.  As equation (40) in \cite{1607.03396[gr-qc]} authors showed that the potential function is of the reflection symmetry and rotation invariant power law form $v(\phi,\sigma)=v_0(\phi^2-\sigma^2)+\mu_0$ by Noether Symmetry approach, so we study this case separately.
\subsubsection{m=2 and n=2 i.e. the potential function $v(\phi,\sigma)=v_0(\phi^2-\sigma^2)+\mu_0$.}
In this case the autonomous system (\ref{auto1a}-\ref{auto1e}) takes the form as follows
\begin{eqnarray}
\dot{H}&=\frac{1}{2}(-r^2+s^2)  \label{auto2a}\\
\dot{y}&=r  \label{auto2b}\\
\dot{r}&=-3Hr-2v_0y  \label{auto2c}\\
\dot{z}&=s  \label{auto2d}\\
\dot{s}&=-3Hs-2v_0z  \label{auto2e}
\end{eqnarray}
The Jacobian matrix for the above system is
\[
J=
\left[ {\begin{array}{ccccc}
	0 & 0 & -r & 0 & s  \\
	0 & 0 & 1 & 0 & 0 \\
	-3r & -2v_0 & -3H & 0 & 0\\
	0 & 0 & 0 & 0 & 1 \\
	-3s & 0 & 0 & -2v_0 & -3H
	\end{array} } \right]
\]

The Jacobian Matrix evaluated at the critical points $(H_c,0,0,0,0)$ (subscript c stands for critical point) takes the form as follows
\[
J(H_c,0,0,0,0)=
\left[ {\begin{array}{ccccc}
	0 & 0 & 0 & 0 & 0  \\
	0 & 0 & 1 & 0 & 0 \\
	0 & -2v_0 & -3H_c & 0 & 0\\
	0 & 0 & 0 & 0 & 1 \\
	0 & 0 & 0 & -2v_0 & -3H_c
	\end{array} } \right]
\]
The eigenvalues of $J(H_c,0,0,0,0)$  are $\lbrace0, \alpha, \beta, \alpha, \beta\rbrace$ where $\alpha=\frac{-3H_c}{2}+ \frac{\sqrt{9H_c^2-8v_0}}{2}$ and $\beta=\frac{-3H_c}{2}- \frac{\sqrt{9H_c^2-8v_0}}{2}$.  The Jordan form of this matrix is
\[
J_{Jordan}(H_c,0,0,0,0)=
\left[ {\begin{array}{ccccc}
	0 & 0 & 0 & 0 & 0  \\
	0 & \alpha & 0 & 0 & 0 \\
	0 & 0 & \beta & 0 & 0\\
	0 & 0 & 0 & \alpha & 0 \\
	0 & 0 & 0 & 0 & \beta
	\end{array} } \right]
\]
The line of non-hyperbolic critical points $(H_c,0,0,0,0)$ are normally hyperbolic \cite{978-94-017-0327-7},\cite{gr-qc/1810.03816}.  The stability of  normally hyperbolic set can be completely classified by considering the sign of the eigenvalues in the remaining directions.
\paragraph{$v_0 \neq 0$.\\ }
First we consider the case $9H_c^2-8v_0=0$.  Then the system (\ref{auto2a}-\ref{auto2e}) has 4-dimensional (4D) stable manifold if $H_c>0$ and 4D unstable manifold if $H_c<0$ (table \ref{table1}).
\begin{table} 
	\caption{ Stability analysis ($v_0 \neq 0$, $9H_c^2-8v_0=0$ )}
	\begin{ruledtabular}
		\begin{tabular}{ c|cc }
			critical Points  & $H_c \neq 0$ ( as $v_0 \neq 0$ )  & stability\\
			\hline
			\multirow{2}{*}{$(H_c,0,0,0,0)$} & $H_c > 0$ & stable node (4-dimensional)\\ & $H_c < 0$ & saddle node (4-dimensional)\\ 
		\end{tabular}\label{table1}
	\end{ruledtabular}
\end{table}

Secondly, when $9H_c^2-8v_0<0$, the vector field on the (y, r)-plane and (z, s)-plane near the critical points behaves like stable focus when $H_c>0$ and unstable focus when $H_c<0$ ( table \ref{table2}).
\begin{table} 
	\caption{ Stability analysis ($v_0 \neq 0$, $9H_c^2-8v_0<0$ )}
	\begin{ruledtabular}
		\begin{tabular}{ c|cc }
			critical Points  & $H_c$  & stability\\
			\hline
			\multirow{3}{*}{$(H_c,0,0,0,0)$} & $H_c > 0$ & stable focus (2-dimensional)\\ & $H_c < 0$ & unstable focus (2-dimensional)\\ & $H_c=0$, $v_0>0$ & center (2-dimensional)
		\end{tabular}\label{table2}
	\end{ruledtabular}
\end{table}

When $H_c=0$,  then $\alpha=\sqrt{-2v_0}$ and $\beta=-\sqrt{-2v_0}$.  The critical points are saddle type for $v_0<0$ and the origin is a center on H=0 hypersurface for $v_0>0$.\\
We next consider $9H_c^2-8v_0>0$ to define the transformation of basis by the matrix P as the following.

\[
P=
\left[ {\begin{array}{ccccc}
	1 & 0 & 0 & 0 & 0  \\
	0 & 1 & 1 & 0 & 0 \\
	0 & \alpha & \beta & 0 & 0\\
	0 & 0 & 0 & 1 & 1 \\
	0 & 0 & 0 & \alpha & \beta
	\end{array} } \right]
\]
and the new  system of equations takes the form as
\begin{widetext}
\begin{eqnarray}
\dot{H}&=-\frac{1}{2}(\alpha Y + \beta R)^2 + \frac{1}{2}(\alpha Z + \beta S)^2\\
\dot{Y}&=\frac{1}{\beta - \alpha}[Y (\alpha \beta + 2v_0)+3\alpha H Y + R(\beta^2 +2v_0) + 3\beta H R]\\
\dot{R}&=-\frac{1}{\beta - \alpha}[R(\alpha \beta + 2v_0)+3\beta H R + Y(\alpha^2 +2v_0)+ 3\alpha H Y]\\
\dot{Z}&=\frac{1}{\beta - \alpha}[Z (\alpha \beta + 2v_0)+3\alpha H Z + S(\beta^2 +2v_0) + 3\beta H S]\\
\dot{S}&=-\frac{1}{\beta - \alpha}[S(\alpha \beta + 2v_0)+3\beta H S + Z(\alpha^2 +2v_0)+ 3\alpha H Z]
\label{JorAuto2}
\end{eqnarray}
\end{widetext}
The orientation of the vector fields of the new system remains same as the original system as $det(P)>0$.  The critical for the system is $(H_c,0,0,0,0)$ which satisfy the following equations 
\begin{eqnarray}
\frac{1}{\beta - \alpha}(\alpha \beta + 2v_0 + 3\alpha H_c)&= \alpha\\
-\frac{1}{\beta - \alpha}(\alpha \beta + 2v_0 + 3\beta H_c)&= \beta\\
\alpha^2+2v_0+3\alpha H_c&=0\\
\beta^2+2v_0+3\beta H_c&=0
\label{ConCri}
\end{eqnarray}
By Hartman-Grobman theorem, the flow along the vectors $[\ 0\ 1\ \alpha\ 0\ 0\ ]^T$ and $[\ 0\ 0\ 0\ 1\ \alpha\ ]^T$ is stable (unstable) near the origin in the shifted coordinate system when $\alpha < 0\ (\alpha >0)$.  Similarly, the flow along the vectors $[\ 0\ 1\ \beta\ 0\ 0\ ]^T$ and $[\ 0\ 0\ 0\ 1\ \beta\ ]^T$ are stable (unstable) when $\beta< 0\ (\beta >0)$ (table \ref{table3}).  As H-axis is the line of critical points, so no flow along the $[\ 1\ 0\ 0\ 0\ 0\ ]^T$- i.e., $\dot{H}=0$.  We find Center Manifold \cite{1706.09284[math.AP], *604.01128[math.AP], *1407.7942[math.CA]} of the above system by shifting the critical points $(H_c,0,0,0,0)$ to the origin.  The Center Manifold at the origin is Y=R=Z=S=0 which gives $\dot{H}=0$.\\

\begin{table} 
	\caption{ Stability analysis (for $v_0 \neq 0$, $9H_c^2-8v_0>0$)}
	\begin{ruledtabular}
		\begin{tabular}{ c|cc }
			critical Points  & $\alpha$ , $\beta$ ($\neq 0$)  & stability\\
			\hline
			\multirow{3}{*}{$(H_c,0,0,0,0)$} & $\alpha<0$ , $\beta<0$ & stable node (4-dim.)\\ & $\alpha <0$ (or $>0$), $\beta>0$(or $<0$) & saddle node (4-dim.)\\ & $\alpha>0$ , $\beta>0$ & unstable node (4-dim.) 		
		\end{tabular}\label{table3}
	\end{ruledtabular}
\end{table}
\paragraph{$v_0 = 0$.\\}
Now we consider the case when the arbitrary constant $v_0$ takes the value 0.  In this case we get three sub cases as follows.
\begin{itemize}
	\item $H_c=0$, $v_0=0$ implies $\alpha=0$ and $\beta=0$.  In this sub case the flow is undetermined analytically.  But numerically we can plot the vector fields.  First we plot the vector fields on the y-r plane as in figure \ref{fig:y_vs_r}.
\begin{figure}[!]
	\includegraphics[scale=0.3]{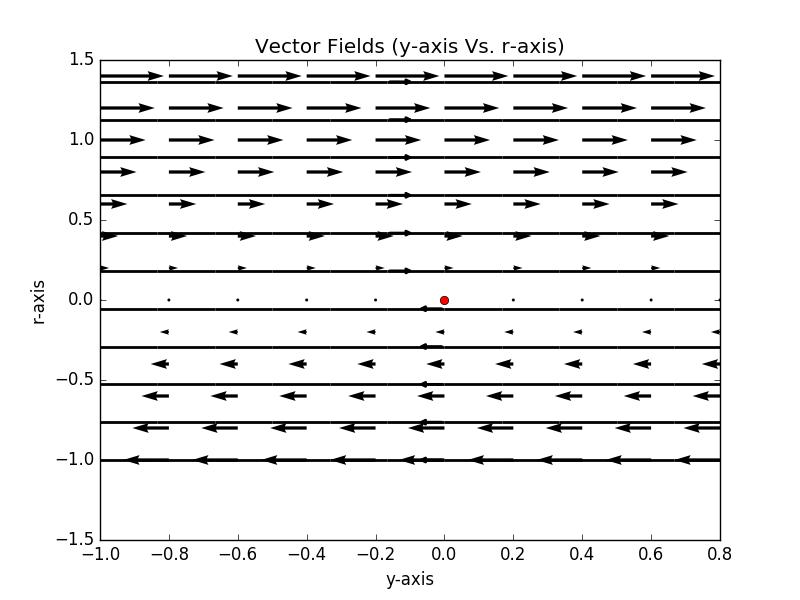}
	\caption{\label{fig:y_vs_r} $H_c=0$ and $v_0=0$.}
\end{figure}
The vector fields for z-axis vs s-axis is exactly same as figure \ref{fig:y_vs_r}.

    \item $H_c>0$, $v_0=0$ implies $-3H_c<0$.  In this sub case, the system (\ref{auto2a})-(\ref{auto2e}) reduces to
\begin{eqnarray}
\dot{H}&=-\frac{1}{2}r^2+\frac{1}{2}s^2  \label{auto1a0}\\
\dot{y}&=r  \label{auto1b0}\\
\dot{r}&=-3Hr  \label{auto1c0}\\
\dot{z}&=s  \label{auto1d0}\\
\dot{s}&=-3Hs  \label{auto1e0}
\end{eqnarray}
In this system the critical points are $(H_c,y_c,0,z_c,0)$ where $H_c, y_c, z_c \in \mathbb{R}$.  So, no flow or vector fields along the eigenvectors correspond to the zero eigenvalue as they are line of critical points.  This argument also gets support in terms of center manifold theory and the center manifold is r=s=0 which indicates $\dot{H}=\dot{y}=\dot{z}=0$.  The flow along the eigenvectors $[\ 0\ -\frac{1}{3H_c}\ 1\ 0\ 0\ ]^T$ and $[\ 0\ 0\ 0\  -\frac{1}{3H_c}\ 1\ ]^T$  are attracting to the CP (table \ref{table4}).

	\item $H_c<0$, $v_0=0$ implies $-3H_c>0$.  In this sub case, The flow is same as in sub case 2 only the flow along the eigenvectors  $[\ 0\ -\frac{1}{3H_c}\ 1\ 0\ 0\ ]^T$ and $[\ 0\ 0\ 0\  -\frac{1}{3H_c}\ 1\ ]^T$  are repelling from the CP (table \ref{table4}).
\end{itemize}

For the special case $v_0=\frac{1}{2}$, the system (\ref{auto2a}-\ref{auto2e}) has two periodic orbits $\Gamma_1(t)=(0,sin\ (t),cos\ (t),0,0)$ and $\Gamma_2(t)=(0,0,0,sin\ (t),cos\ (t))$ around the origin on the (y-r)-plane and (z-s)-plane respectively.  Now we write the system (\ref{auto2a}-\ref{auto2e}) as $\dot{\Upsilon}=f(\Upsilon)$ where $\Upsilon=(H,y,r,z,s)^T$ and $f\in C^{\infty}(\mathbb{R}^5)$.  To find the stability of the periodic orbit we need to find the value of $\int_{0}^{2\pi} \nabla f(\Gamma_1(t)) dt$ where $2\pi$ is the period of $\Gamma_1(t)$.  We find $\int_{0}^{2\pi} \nabla f(\Gamma_1(t)) dt=0$.  So $\Gamma_1(t)$ (similarly $\Gamma_2(t)$) belongs to a continuous band of cycles.
\begin{table} 
	\caption{ Stability analysis (for $v_0=0$)}
	\begin{ruledtabular}
		\begin{tabular}{ c|cc }
			critical Points  & $H_c$  & stability\\
			\hline
			$(H_c,0,0,0,0)$ & $=0$ & unstable (saddle) (4-dimensional)\\
			\hline
			\multirow{2}{*}{$(H_c,y_c,0,z_c,0)$}& $>0$ & stable node (4-dimensional)\\ & $<0$ & unstable node (4-dimensional) 		
		\end{tabular}\label{table4}
	\end{ruledtabular}
\end{table}

\subsubsection{ $m>2$ and $n>2$}
The Jacobian matrix evaluated at the critical points $(H_c,0,0,0,0)$ is
\[
J(H_c,0,0,0,0)=
\left[ {\begin{array}{ccccc}
	0 & 0 & 0 & 0 & 0  \\
	0 & 0 & 1 & 0 & 0 \\
	0 & 0 & -3H_c & 0 & 0\\
	0 & 0 & 0 & 0 & 1 \\
	0 & 0 & 0 & 0 & -3H_c 
	\end{array} } \right]
\]

\begin{figure}[!]
	\includegraphics[scale=0.4]{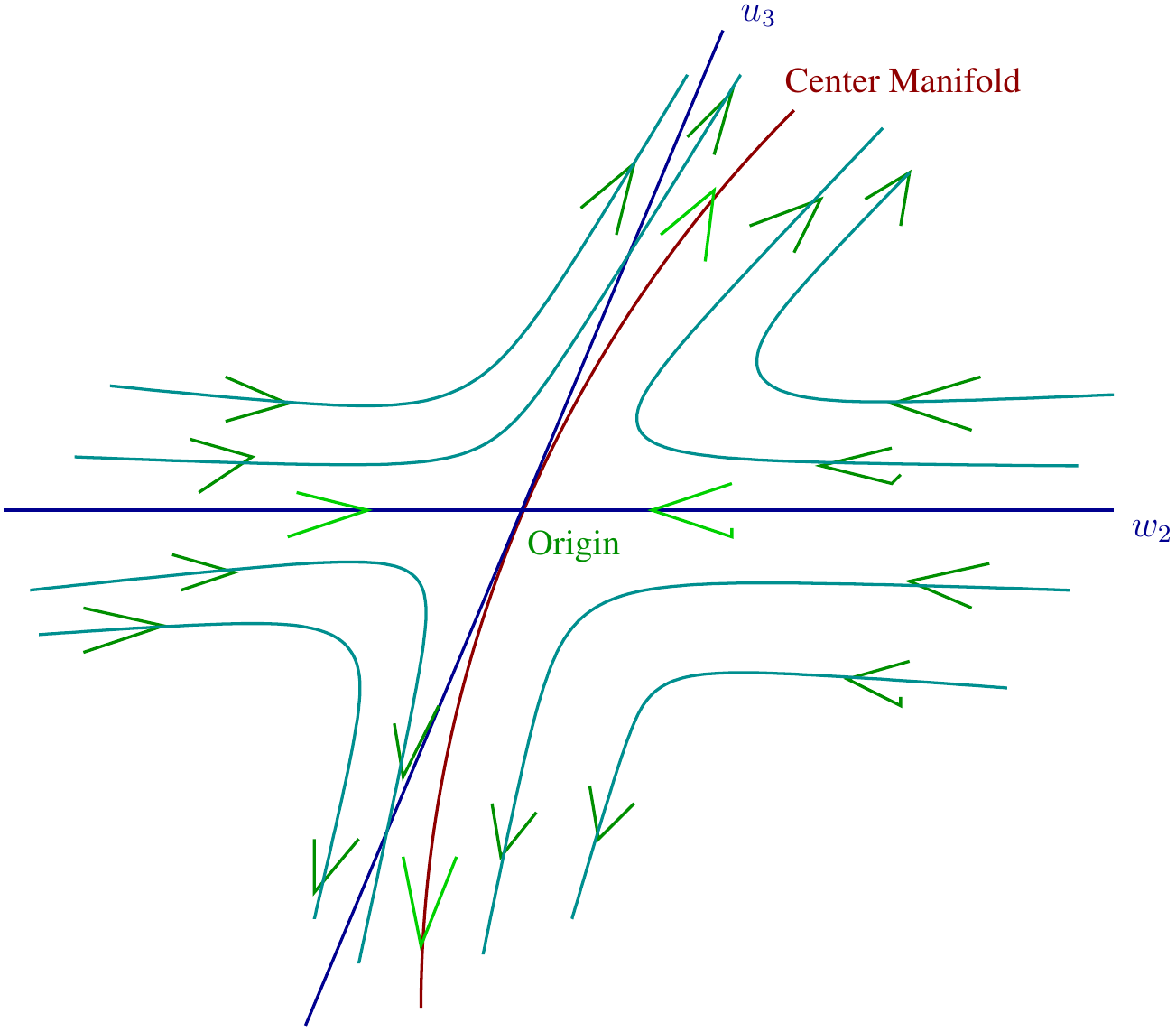}
	\caption{\label{fig:cenman1} For $H_c<0$ and m even.}
\end{figure}

\begin{figure}[!]
	\includegraphics[scale=0.4]{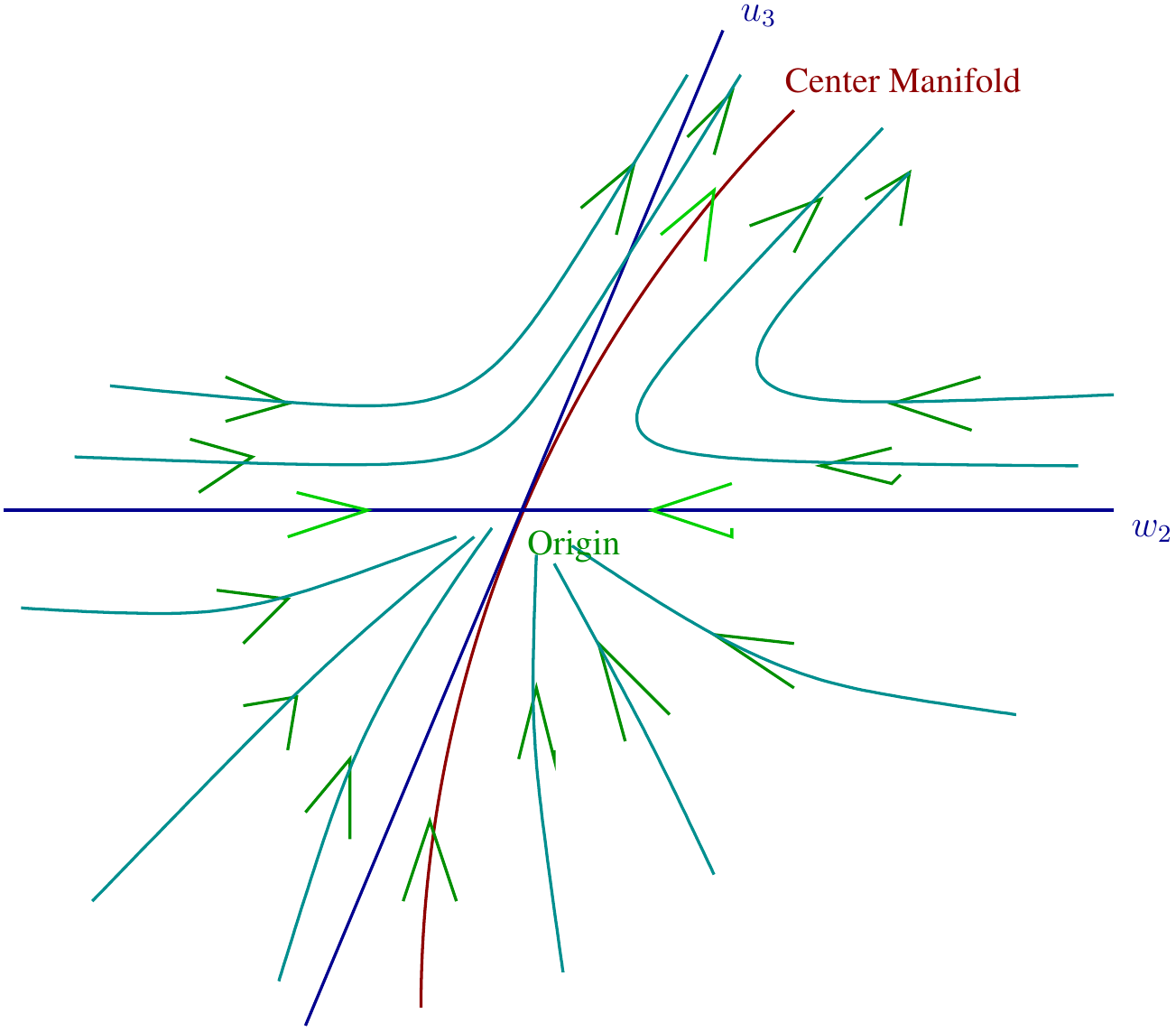}
	\caption{\label{fig:cenman2} For $H_c<0$ and m odd.}
\end{figure}

\begin{figure}[!]
	\includegraphics[scale=0.4]{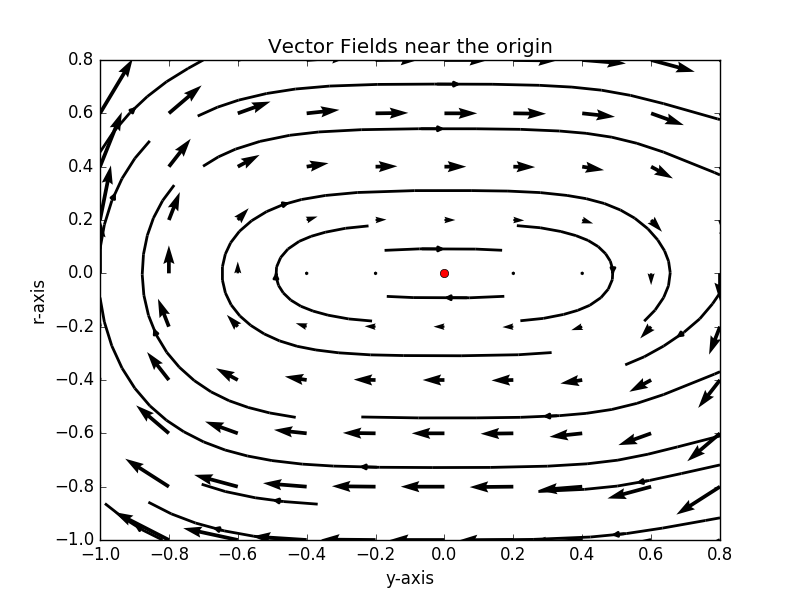}
	\caption{\label{fig:vector1} For n even, origin is a center (for equation \ref{PjtAuto1b} and \ref{PjtAuto1c}).}
\end{figure}

\begin{figure}[!]
	\includegraphics[scale=0.4]{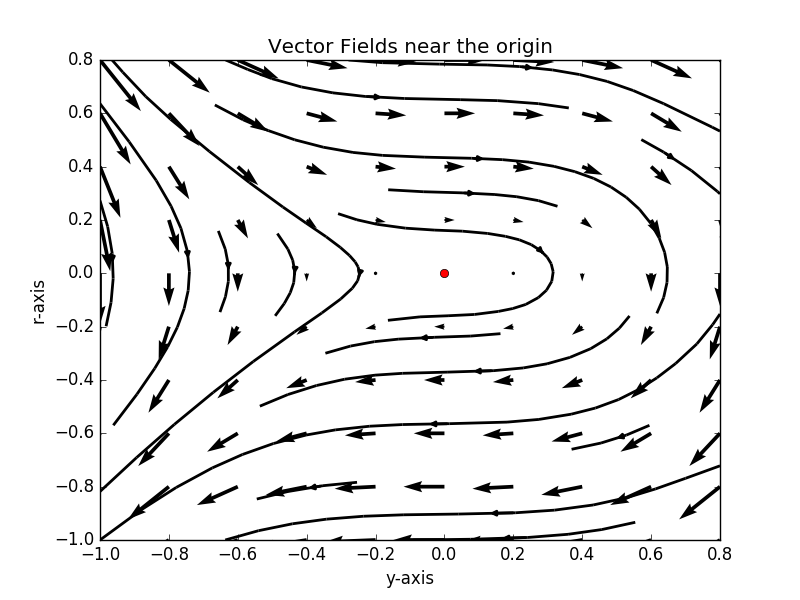}
	\caption{\label{fig:vector2} For n odd, origin is a cusp (for equation \ref{PjtAuto1b} and \ref{PjtAuto1c}).}
\end{figure}

The algebraic as well as geometric multiplicity of the eigenvalues 0 and $-3H_c$ are three and two respectively.  The eigenvectors corresponding to the eigenvalue 0 are $u_1=[\ 1\ 0\ 0\ 0\ 0\ ]^T$, $u_2=[\ 0\ 1\ 0\ 0\ 0\ ]^T$ and $u_3=[\ 0\ 0\ 0\ 1\ 0\ ]^T$.  The eigenvectors corresponding to the eigenvalue $-3H_c$ are $w_1=[\ 0\ -\frac{1}{3H_c}\ 1\ 0\ 0\ ]^T$ and $w_2=[\ 0\ 0\ 0\ -\frac{1}{3H_c}\ 1\ ]^T$.  By Hartman-Grobman Theorem, the flow along $[\ 0\ -\frac{1}{3H_c}\ 1\ 0\ 0\ ]^T$ and $[\ 0\ 0\ 0\ -\frac{1}{3H_c}\ 1\ ]^T$ near the critical points $(H_c,0,0,0,0)$ are stable (unstable) if $-3H_c<0$ $(-3H_c>0)$.  To find the flow along the eigenvectors of 0, we have to use Center Manifold Theory.  The center manifold is as follows
\begin{equation}
r=-\frac{nv_0}{3H_c}y^{n-1} + higher\ degree\ terms.
\label{cenman1}
\end{equation}
\begin{equation}
s=-\frac{nv_0}{3H_c}z^{m-1} + higher\ degree\ terms.
\label{cenman2}
\end{equation}
The center manifold (\ref{cenman1}) is tangent to $[\ 0\ 1\ 0\ 0\ 0\ ]^T$ near the origin and the flow along the center manifold is determined by 
\begin{equation}
\dot{y}=-\frac{nv_0}{3H_c}y^{n-1} + higher\ degree\ terms.
\label{flow1}
\end{equation}
Similarly, the flow near the origin along the center manifold (\ref{cenman2}) is determined by 
\begin{equation}
\dot{z}=-\frac{nv_0}{3H_c}z^{m-1} + higher\ degree\ terms.
\label{flow2}
\end{equation}
So the flow near the origin (after shifting the critical point to origin) is saddle for m (or n) is even and $H_c<0$ as figure (\ref{fig:cenman1}).  On the other hand, the flow near the origin is saddle-node for m (or n) is odd and $H_c<0$ as figure (\ref{fig:cenman2}).  We get the stable node near origin for $H_c>0$ and m being even positive integer and saddle for $H_c>0$ and m being odd.\\
As $[\ 1\ 0\ 0\ 0\ 0\ ]^T$ is line of CPs, so there is no flow or vector fields along $[\ 1\ 0\ 0\ 0\ 0\ ]^T$ near the CPs. So for $H_c=0$ we analyze the behavior of the vector fields on the hypersurface   H=0 (name it $\mathscr{H}$).  Now on $\mathscr{H}$ the equations (\ref{auto1b}) and (\ref{auto1c}) reduce to
\begin{eqnarray}
\dot{y}&=r  \label{PjtAuto1b}\\
\dot{r}&=-nv_0y^{n-1}  \label{PjtAuto1c}
\end{eqnarray}
Similarly, on $\mathscr{H}$ the equations (\ref{auto1d}) and (\ref{auto1e}) reduce to
\begin{eqnarray}
\dot{z}&=s  \label{PjtAuto1d}\\
\dot{s}&=-nv_0z^{m-1}  \label{PjtAuto1e}
\end{eqnarray}
On $\mathscr{H}$ the system of equations (\ref{PjtAuto1b}) and (\ref{PjtAuto1c}) are uncouple with (\ref{PjtAuto1d}) and (\ref{PjtAuto1e}).  So we analyze them independently.  In the equations (\ref{PjtAuto1b}) and (\ref{PjtAuto1c}), the power of y namely $n-1 \geqslant 2$ and origin ($\mathscr{O}$) is only critical point.  First we choose n is even.  So n-1 is odd, say n-1=2k+1, for some $k \geqslant 1$.  So the origin is a focus or a center for $(-nv_0)<0$ i.e $v_0>0$ (for reference, see theorem 2 and 3 in section 2.11 in \cite{978-1-4613-0003-8}) where numerical computations ensure that $\mathscr{O}$ is a center (figure \ref{fig:vector1}) and for all $\epsilon >0$ there exists a $\delta >0$ such that for all $x \in \mathscr{N}_{\delta}(\mathscr{O})$ and $t \geqslant 0$ we have $\phi_t(x)\in \mathscr{N}_\epsilon(\mathscr{O})$.  So $\mathscr{O}$ is stable.  On the other hand, $\mathscr{O}$ is a (topological) saddle for $v_0<0$.  So $\mathscr{O}$ is unstable in this case.  If n is odd, then n-1=2k  for some $k \geqslant 1$.  In this case the origin is a cusp \cite{978-1-4613-0003-8} as in figure (\ref{fig:vector2}).   Similarly, replacing n by m, we get the same stability criteria at the origin for the equations (\ref{PjtAuto1d}) and (\ref{PjtAuto1e}) projecting on $\mathscr{H}$ (table \ref{table5}).
\begin{table} 
	\caption{ Stability analysis (for $m>2$ and $n>2$)}
	\begin{ruledtabular}
		\begin{tabular}{ c|cc }
			critical Points  & $H_c$  & stability\\
			\hline
			m=even, n=even & $H_c>0$, $v_0>0$ & stable node (4-dimensional)\\
			\hline
			m=odd, n=odd & $H_c>0$, $v_0>0$ & saddle node (4-dimensional)\\	
			\hline
			m=odd, n=even \\or\\ m=even, n=odd & $H_c>0$, $v_0>0$ & saddle (4-dimensional)\\	
		\end{tabular}\label{table5}
	\end{ruledtabular}
\end{table}


\subsubsection{n=2 (or $n>2$) and $m>2$ (or m=2)}
For this sub case we can use the above two sub cases to analyze the phase-space.

\subsubsection{Bifurcation Analysis}
\begin{figure}
	\includegraphics[scale=0.4]{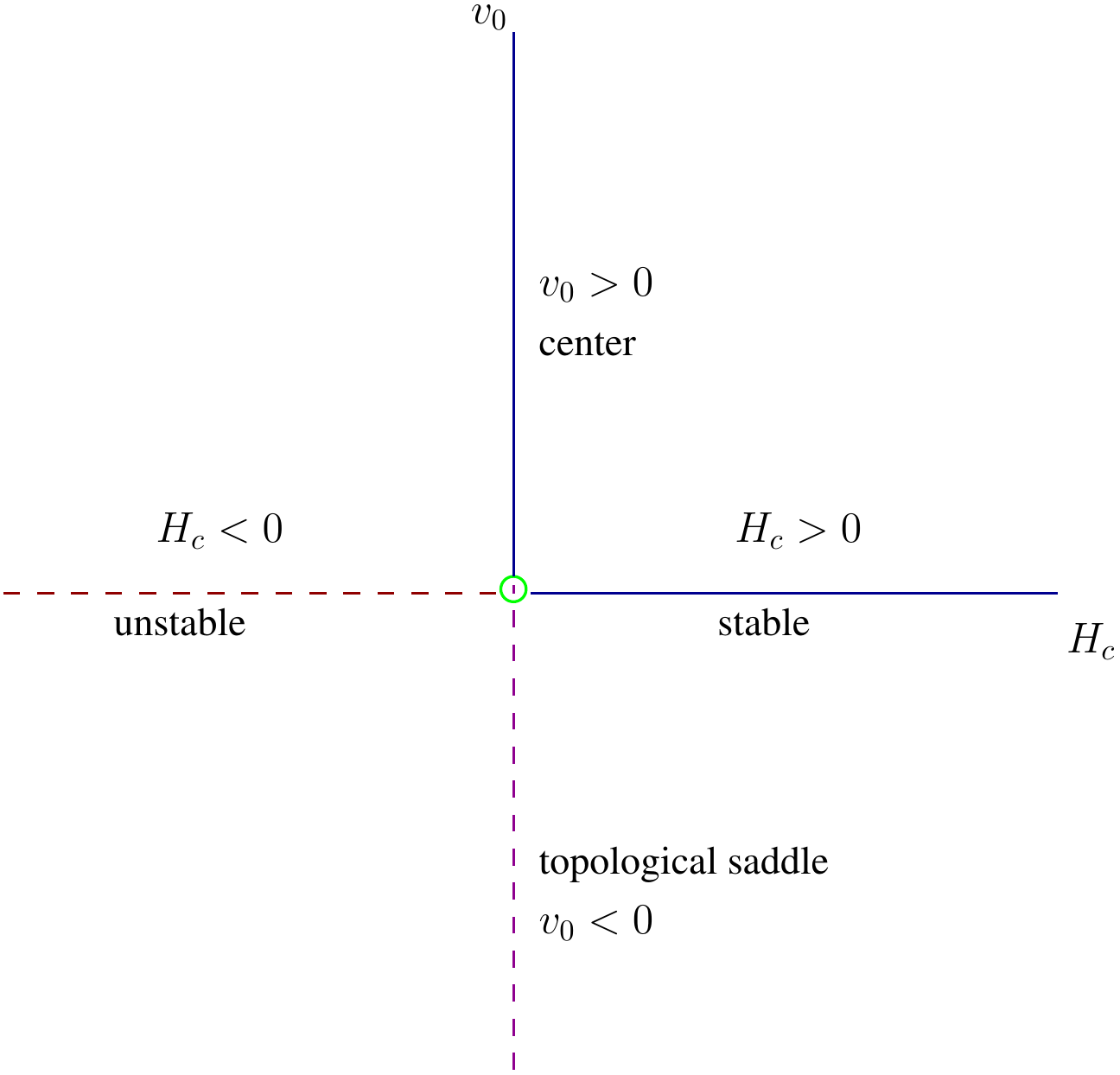}
	\caption{\label{fig:bifurcation} Bifurcation diagram for $v_0=0$.}
\end{figure}

For $v_0=0$, on the eigenspace of $-3H_c$, the vector fields are attracting towards the CPs for $H_c>0$ and repelling for $H_c<0$.  At $H_c=0$, we get phase portrait as in figure \ref{fig:y_vs_r}.  So the line of CPs r=0 is unstable.  Thus at $H_c=0$, the system is structurally unstable as small perturbation at $H_c=0$, we get different characteristics of the vector fields.  So, taking $H_c$ as a parameter, the bifurcation value is $H_c=0$ and bifurcation point is the origin \cite{1807.09525[math.DS], *gr-qc/0307044}. \\
At $H_c=0$, for n, m are even and $v_0>0$, the origin is a focus or center.  On the other hand, for $v_0<0$, the origin is a (topological) saddle.  For $v_0=0$, the vector fields near origin discussed above (figure \ref{fig:y_vs_r}).  So $v_0=0$ is a bifurcation value (figure \ref{fig:bifurcation}).

\subsection{ $v(\phi,\sigma)=e^{-\lambda(\phi + \sigma)}$ }
We next choose the potential function $v(\phi,\sigma)=e^{-\lambda(\phi + \sigma)}=e^{\kappa(\phi + \sigma)}$ (here $-\lambda=\kappa$, say).  Here $\lambda$ is a dimensionless constant characterizing the slop of the potential for $\phi$ and $\sigma$.  Further we assume $\lambda\geqslant 0$ since we can make them positive through $\phi \rightarrow -\phi$ and $\sigma \rightarrow -\sigma$ if some of them are negative.   Now we choose $\dot{\phi}=r$ and $\dot{\sigma}=s$.  So we get the system as follows:
\begin{eqnarray}
\dot{H}&=\frac{1}{2}(-r^2+s^2)  \label{auto3a}\\
\dot{r}&=-3Hr+\kappa v  \label{auto3b}\\
\dot{s}&=-3Hs-\kappa v  \label{auto3c}\\
\dot{v}&=\kappa v(r+s)  \label{auto3d}
\end{eqnarray}

The Jacobian matrix of the system is
\[
J=
\left[ {\begin{array}{cccc}
	0 & -r & s & 0  \\
	-3r & -3H & 0 & \kappa \\
	-3s & 0 & -3H & -\kappa\\
	0 & \kappa v & \kappa v & \kappa(r+s) 
	\end{array} } \right]
\]
The critical points of the system are $(H_c,r_c,s_c,v_c)$ where $r_c=-s_c$ and $3H_cr_c=\kappa v_c \neq 0$ $(H_c,r_c,s_c,v_c \in \mathbb{R})$.  The Jacobian matrix evaluated at the critical points is

\[
J(H_c,r_c,-r_c,v_c)=
\left[ {\begin{array}{cccc}
		0 & -r_c & -r_c & 0  \\
		-3r_c & -3H_c & 0 & \kappa \\
		3r_c & 0 & -3H_c & -\kappa\\
		0 & \kappa v_c & \kappa v_c & 0 
	\end{array} } \right]
\]

The eigenvalues are $ \lbrace 0,0, -3H_c, -3H_c \rbrace $.  The Jordan form of $J(H_c,r_c,-r_c,v_c)$ is 
\[
J_{Jordan}(H_c,r_c,-r_c,v_c)=
\left[ {\begin{array}{cccc}
		0 & 0 & 0 & 0  \\
		0 & 0 & 0 & 0 \\
		0 & 0 & -3H_c & 1 \\
		0 & 0 & 0 & -3H_c 
	\end{array} } \right]
\]

The change of basis matrix is 

\[
P=
\left[ {\begin{array}{ccccc}
	H_c & \kappa & 0 & r_c  \\
	-r_c & 0 & -(3r_c^2 + \kappa^2 v) & 3H_c \\
	r_c & 0 & (3r_c^2 + \kappa^2 v) & 0 \\
	0 & 3r_c & 0 & -\kappa v_c 
	\end{array} } \right]
\]

By this (P matrix) change of basis, the system (\ref{auto3a}-\ref{auto3d}) takes the form

\begin{widetext}
\begin{eqnarray}
\dot{\varmathbb{H}}&=(-\frac{9H_c}{2}+ \frac{3r_c}{H_c}(r_c + \kappa H_c))V^2 + 3(3r_c^2+\kappa^2 v_c)SV + \frac{3\kappa r_c}{H_c}RV + 3(2y_c + \kappa H_c)\varmathbb{H}V\\
\dot{R}&=(-3H_c^2)\varmathbb{H}V + -3H_c(r_c + H_c)V^2\\
\dot{S}&=-\frac{3H_cr_c}{3r_c^2+\kappa^2 v_c}\varmathbb{H}^2 + \frac{3(3+3\kappa^2)H_cr_c}{2(3r_c^2 +\kappa^2 v_c)}V^2 - \frac{9r_c^2 + \kappa^2 v_c}{3r_c^2 + \kappa^2 v_c}\varmathbb{H}V - (3H_c)\varmathbb{H}S  -\frac{3\kappa r_c}{3r_c^2 + \kappa^2 v_c}\varmathbb{H}R - (6r_c)SV \nonumber \\  &+ (-\frac{\kappa}{H_c}+ \frac{3\kappa^2 r_c}{3r_c^2+\kappa^2 v_c})RV - (3\kappa)RS - \frac{3r_c\kappa}{3r_c^2+\kappa^2 v_c}R + \frac{\kappa^2v_c}{3r_c^2+\kappa^2 v_c}V\\
\dot{V}&= (-3H_c)\varmathbb{H}V + (-3\kappa)RV + (-3r_c)V^2
\label{JorAuto3}
\end{eqnarray}
\end{widetext}

\subsubsection{Stability Analysis}
The orientation of the vector fields of new autonomous system is same as the original one if $r_c<0$ and reverse if $r_c>0$.  The critical points $(H_c, r_c, -r_c, v_c)$ changes to $(0,\frac{v_c}{3r_c}, -\frac{r_c}{3r_c^2+\kappa^2 v_c}, 0)$ where $3H_cr_c= \kappa v_c$.  The flow along the vectors $[\ 0\ -(3r_c^2+\kappa^2v_c)\ (3r_c^2+\kappa^2v_c)\ 0\ ]^T$ and $[\ r_c\ 3H_c\ 0\ -\kappa v_c\ ]^T$ is stable (unstable) when $H_c>0\ (H_c<0)$.  The flow along the eigenvectors corresponding to the eigenvalue 0 can not be determined using Hartman-Grobman theorem.  So we use Center Manifold Theory to find it.  The Center Manifold at the origin is 
\begin{eqnarray}
	S&=-\frac{r_c}{3r_c^2+\kappa^2v_c}\varmathbb{H}^2-\frac{\kappa r_c}{H_c(3r_c^2+\kappa^2 v_c)}\varmathbb{H}R  \nonumber \\
	 & + Higher\ degree\ terms.\label{cenman3} \\
	V&=0 \label{cenman4}	 
\end{eqnarray}
These result $\dot{\varmathbb{H}}=0$ and $\dot{R}=0$.  Thus no flow along $[\ H_c\ -r_c\ r_c\ 0\ ]^T$ and $[\ \kappa\ 0\ 0\ 3r_c\ ]^T$ near the origin in the shifted coordinate system ($(0,\frac{v_c}{3r_c}, -\frac{r_c}{3r_c^2+\kappa^2 v_c}, 0)$ to the origin).\\

\subsubsection{Bifurcation Analysis}
The local dynamics of a critical point may depends one or more arbitrary parameters and a subtle continuous change of parameter results dramatic change in the dynamics when the system passes through a structural instability or the parameter of the system crosses the bifurcation value \cite{1810.10120[math.AP]}.  The system of equations~ (\ref{auto3a})-(\ref{auto3d}) is structurally unstable when $H_c=0$.  Thus taking $r_c$ and $v_c$ fixed, the values of the parameter $\kappa$ for which $H_c=0$ (by the relation $3H_cr_c=\kappa v_c$) are the bifurcation values where origin is the bifurcation point.  So for each fixed $r_c$ and $v_c$ we get different bifurcation values.

\section{Discussion}
The couple scalar field dynamical dark energy model (known as quintom model) has been studied in cosmological perspective in formulation of dynamical system analysis \cite{gr-qc/1810.03816, *1807.05236[gr-qc]}.  The coupled potential of the quintom model in chosen as a linear combination of the power-law of the two scalar fields and an exponential product form of the scalar fields.  For the linear combination of the power law form of the potential several cases have been discussed for different choices of the powers.  In most of the cases, there is a line of critical points: $(H_c, 0, 0, 0, 0)$ with $H_c$ is the value of Hubble parameter when $\dot{H}=0$.  The center manifold is characterized by $\dot{H}=0$ when powers $(m, n)$ are chosen to be 2.  When $m>2$, $n>2$, the center manifold is determined by equations (\ref{cenman1}) and (\ref{cenman2}) and the flow along the center manifold are given by equations (\ref{flow1}) and (\ref{flow2}).  It is found that $v_0=0$ is a bifurcation point but it is not interesting as coupled potential is zero.  However, it has been shown that for $v_0>0$, the critical point is a focus or center.\par
On the other hand, for the exponential product form of choice of the potential, the non-hyperbolic critical point is characterized by center manifold given by equations (\ref{cenman3}) and (\ref{cenman4}) and it is found that the system is structurally unstable for $H_c=0$ and it corresponds to a bifurcation point.\par
Finally, from cosmological point of view, the critical points of the present quintom model can be analysed as follows:\\  The line of critical points $(H_c,0,0,0,0)$ represents the phantom barrier in the cosmological context as $w_{eff}=-1$ and $q=1$ at this critical point.  So as expected it behaves as phantom field evolution.  In the autonomous system (\ref{auto1a0})-(\ref{auto1e0}), for the critical point $(H_c,y_c,0,z_c,0)$ one gets  $w_{eff}=-1$ and $q=-1$.  Thus the quintom model describes cosmic evolution with a cosmological term- i.e., the model describes the $\Lambda$CDM era of evolution.  Similar cosmic evolution can be obtained for the critical point $(H_c,r_c,s_c,\nu_c)$ for the autonomous system (\ref{auto3a})-(\ref{auto3d}).  Therefore, from the dynamical system analysis of the present quintom model one may conclude that the present quintom model mostly describes the $\Lambda$CDM phase of cosmic evolution.

\section{Acknowledgments}
The author S. Mishra is grateful to CSIR, Govt. of India for giving Junior Research Fellowship (CSIR Award No: 09/096 (0890)/2017- EMR - I) for the Ph.D work. \\

\bibliography{quintom}
\end{document}